\documentclass[
aps,
prl,
reprint,
superscriptaddress,
longbibliography
]{revtex4-2}

\usepackage{graphicx}
\usepackage{amsmath,amssymb}
\usepackage{hyperref}

\usepackage{graphicx}% Include figure files
\usepackage{dcolumn}% Align table columns on decimal point
\usepackage{bm}% bold math
\usepackage{xcolor}
\usepackage[normalem]{ulem}
\definecolor{RevisionRed}{RGB}{180,0,0}
\definecolor{DeleteBlue}{RGB}{0,0,180}

\newcommand{\del}[1]{\textcolor{DeleteBlue}{\sout{#1}}}
\begin{document}

%\linenumbers

\preprint{APS/123-QED}

\title{{Experimental demonstration of asynchronous measurement-device-independent quantum cryptographic conferencing}}% Force line breaks with \\

\author{Haotao Zhu}
\affiliation{School of Electrical \& Electronic Engineering, Nanyang Technological University, Singapore 639798, Singapore}
\affiliation{Quantum Science and Engineering Centre (QSec), Nanyang Technological University, Singapore 639798, Singapore}
\affiliation{Division of Physics and Applied Physics, School of Physical and Mathematical Sciences, Nanyang Technological University, Singapore 637371, Singapore}

\author{Zhenhua Li}
\affiliation{China Telecom Research Institute, Beijing 102209, China}

\author{Shuai Zhao}
\affiliation{School of Cyberspace, Hangzhou Dianzi University, Hangzhou 310018, China}

\author{Xiaodan Lyu}
\affiliation{Majulab, International Research Laboratory IRL 3654, CNRS, Université Côte d’Azur, Sorbonne Université, National University of Singapore, Nanyang Technological University, Singapore, Singapore}

\author{Shihao Ru}
\affiliation{School of Electrical \& Electronic Engineering, Nanyang Technological University, Singapore 639798, Singapore}
\affiliation{Division of Physics and Applied Physics, School of Physical and Mathematical Sciences, Nanyang Technological University, Singapore 637371, Singapore}
\affiliation{National Centre for Advanced Integrated Photonics, Nanyang Technological University, 21 Nanyang Link, Singapore 637371, Singapore}

\author{Yizhi Huang}
\affiliation{China Telecom Quantum Information Technology Group Corporation Limited, Hefei, Anhui 230088, China}

\author{Zitong Xu}
\affiliation{School of Electrical \& Electronic Engineering, Nanyang Technological University, Singapore 639798, Singapore}
\affiliation{Division of Physics and Applied Physics, School of Physical and Mathematical Sciences, Nanyang Technological University, Singapore 637371, Singapore}
\affiliation{Center for Quantum Technologies, Nanyang Technological University, Singapore 637371, Singapore}

\author{Rui Qu}
\affiliation{School of Electrical \& Electronic Engineering, Nanyang Technological University, Singapore 639798, Singapore}
\affiliation{Division of Physics and Applied Physics, School of Physical and Mathematical Sciences, Nanyang Technological University, Singapore 637371, Singapore}
\affiliation{Center for Quantum Technologies, Nanyang Technological University, Singapore 637371, Singapore}

\author{Weibo Gao}
\email{wbgao@ntu.edu.sg}
\affiliation{School of Electrical \& Electronic Engineering, Nanyang Technological University, Singapore 639798, Singapore}
\affiliation{Quantum Science and Engineering Centre (QSec), Nanyang Technological University, Singapore 639798, Singapore}
\affiliation{Division of Physics and Applied Physics, School of Physical and Mathematical Sciences, Nanyang Technological University, Singapore 637371, Singapore}
\affiliation{Center for Quantum Technologies, Nanyang Technological University, Singapore 637371, Singapore}

%\date{\today}

\date{\today}% It is always \today, today,

\begin{abstract}
Quantum network enable a variety of quantum information processing tasks, where multi-user quantum communication is one of the important objectives. Quantum cryptographic conferencing {(QCC)} serves as an essential solution to establish  secure keys to realize secure multi-user communications.
However, existing QCC implementations have been fundamentally limited by the low probability of multi-user coincidence detection to measure or construct the Greenberger-Horne-Zeilinger (GHZ) entangled state. 
In this work, we report the experimental realization of QCC eliminating the need for coincidence detection, where the GHZ state is constructed by correlating detection events occurring within the coherence time, thereby greatly enhancing the success probability of GHZ-state measurement. Meanwhile, to establish and maintain high-visibility GHZ measurement among three independent users, we developed a three-party phase compensation scheme combined with precise temporal and polarization alignment within a time-bin–phase encoding framework.
Furthermore, we designed an efficient pairing strategy to simplify subsequent data processing and enhance  processing efficiency.
Based on these techniques, we successfully performed QCC experiments over total loss  of 
66.3~dB, {with the channel loss of 51.24 dB},
achieving secure key rates of 5.4~bit/s, whereas
previous QCC experiments have been limited to {100 km}.
The results 
establish a new regime of {scalable, }multi-user quantum communication and paving the way for metropolitan quantum networks.
%\begin{description}
%\end{description}
\end{abstract}

\maketitle

%\tableofcontents

\textit{Introduction—}
With rapid development of quantum technology, it is of central interest to establish secure and efficient quantum communication networks~\cite{kimble2008quantum,wehner2018quantum,pittaluga2025long}.
%As progress towards the quantum internet accelerates, the ability to establish secure and efficient quantum networks has become a central goal. 
Quantum key distribution (QKD)~\cite{bennett2014quantum,ekert1991quantum,gisin2002quantum,scarani2009security,xu2020secure} offers an effective method to establish information-theoretic security between two nodes by encoding information into quantum states of photons. It guarantees information-theoretic security based on the laws of quantum mechanics. 
However, in multi-user quantum networks, enabling secure communication among $N$ parties with pairwise QKD requires the preparation of up to $N-1$ copies of the key, which is highly inefficient~\cite{epping2017multi}. 
An effective alternative is quantum cryptographic conferencing (QCC)~\cite{bose1998multiparticle,chen2004multi,murta2020quantum}, where secret keys are distilled directly from multi-party entangled states shared among multiple users. 

\begin{figure*}[t]
	\centering
	\includegraphics[width=0.8\textwidth]{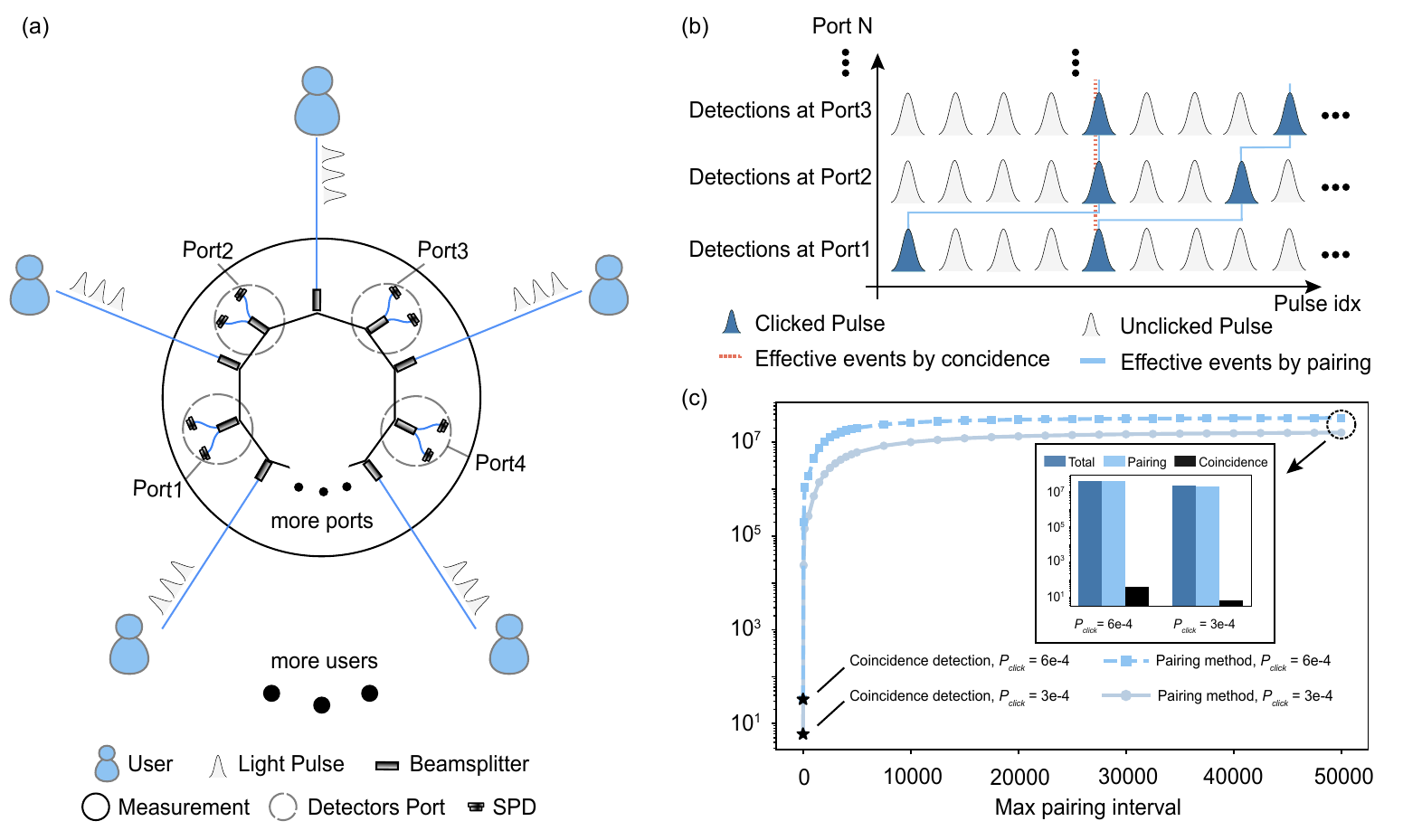}
	\caption{(a) Multiple users send light pulses to a central measurement node, where a ring-structured detection architecture enables pairwise single-photon interference among all users through multiple detector ports. (b) At each detection port, the pulse sequence indicates the responses details of single-photon detectors (SPD).
		Blue pulses indicate time slots where {only} one single-photon detector registers a click {at the port}, whereas gray pulses correspond to no detector response. In the coincidence-based method, an effective event occurs only when all ports produce detections at the same pulse index. In contrast, the proposed pairing method identifies valid events by finding detector responses from each port within the coherence time window, without requiring detections to occur at identical pulse positions. (c) We present the experimental results showing the variation in the number of effective events as a function of the maximum pairing interval under two different pulse detection probabilities, \( P_{\mathrm{click}} \). For comparison, the number of effective events obtained from coincidence detection is also plotted, while {Total} denotes the total number of detected pulses. The {Pairing} results indicate that most detection clicks can be successfully paired, which reflects a transition from $P_{\mathrm{click}}^{N}$ to $P_{\mathrm{click}}$ in event probability.
	}
	\label{fig:mainidea}
\end{figure*}

Within this framework, various QCC protocols have been proposed~\cite{epping2017multi,grasselli2018finite,fu2015long,zhao2020phase,cao2021coherent,carrara2023overcoming,proietti2021experimental,hahn2020anonymous,grasselli2019conference,lu2025repeater,xie2024multi}, which can be broadly categorized into two classes: entanglement-based and time-reversed approaches.
In the entanglement-based scheme, a central node prepares a multi-party entangled state—such as a Greenberger–Horne–Zeilinger (GHZ) state~\cite{greenberger1989going}—and distributes it to multiple users, who then perform local measurements to establish a shared secret key.
The time-reversed class refers to schemes in which the central node performs the measurement, and the GHZ state is effectively created through post-selection of the detection events \cite{pan1998greenberger}. Several experimental demonstrations
following the time-reversed protocols have been reported~\cite{yang2024experimental,du2025experimental}. However, both approaches are fundamentally limited by the intrinsically low probability of multi-photon coincidence detections, making them impractical for scalable scenarios. Although Ref.~\cite{lu2025repeater} proposes an AMDI-QCC protocol without coincidence detections, its experimental realization remains unexplored. The three-user implementation is highly non-trivial, particularly due to the complexity of achieving multi-user time and phase stabilization. It requires precise synchronization, stringent control of interference-induced error rates, and also a sophisticated pairing strategy. More importantly, the effect of the global phase, which is not explicitly considered in Ref.~\cite{lu2025repeater}, is taken into account here and is shown to be crucial in practice.

 Here, we demonstrate three-user AMDI-QCC experiment, 
achieving secure key generation over a total loss of 66.3~dB with a key rate of
\(5.4~\)bit/s, while
previous QCC experiments have been limited to 100 km~\cite{proietti2021experimental,du2025experimental,yang2024experimental}.  We constructed a three-user interference setup with active temporal and polarization feedback to ensure high-visibility interference between every two users. Furthermore, an efficient pairing strategy was designed to rapidly associate interference events, which accelerates valid-pair generation and simplifies the subsequent data processing. Based on this, we analyzed how the phase of each paired event affects the GHZ-state visibility, and accordingly developed a tailored phase-compensation method. As a result, we attain a GHZ-state measurement error of {39.42\%}, close to the {37.5\%} theoretical limit~\cite{fu2015long}. {The secure key rates provide supporting evidence for the expected near-linear O(T) scaling of AMDI-QCC,}
%\add{From the total loss of
%	51.8 to 66.3 dB, the corresponding single-arm transmittance decreases
%	from \(1.88\times 10^{-2}\) to \(6.18\times 10^{-3}\), while the key rate
%	decreases from \(1.64\times 10^{-7}\) to \(3.75\times 10^{-8}\) per pulse.
%	The values of \(R/T\) are \(8.59\times 10^{-6}\) and
%	\(6.08\times 10^{-6}\), respectively, indicating that the measured key-rate
%	scaling is broadly consistent with an \(O(T)\) dependence, where \(T\)
%	denotes the transmittance from a single user to the central measurement node,}
surpassing the conventional key rate scaling of multi-party protocols~~\cite{fu2015long,zhao2020phase}, where the key rate was limited by an $O(T^{N})$ scaling in previous QCC experiments~\cite{yang2024experimental,du2025experimental}.

\begin{figure*}[http]
	\centering
	\includegraphics[width=0.8\textwidth]{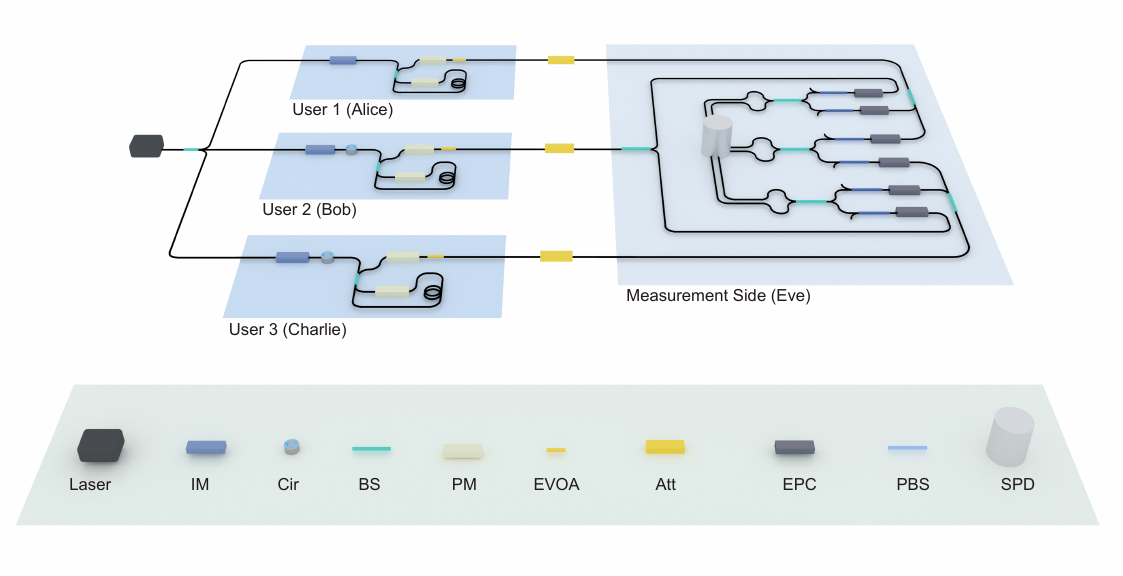}
	\caption{Experimental Setup. A single laser is split into three beams serving as sources for Alice, Bob, and Charlie. Each beam is modulated by an intensity modulator and a Sagnac loop comprising a circulator, a beam splitter, and a phase modulator, followed by discrete 16-level phase modulation for randomization. The pulses are attenuated to the single-photon level using an electronically variable optical attenuator before transmission through a channel implemented by an additional optical attenuator. At the receiver, three BSs enable pairwise interferences between Alice–Bob, Bob–Charlie, and Charlie–Alice. Polarization feedback is applied using electronic polarization controllers and polarization beam splitters, and the outputs are detected by single-photon detectors. Abbreviations: IM, intensity modulator; Cir, circulator; BS, beam splitter; Att, attenuator; PM, phase modulator; EVOA, electronically variable optical attenuator; EPC, electronic polarization controller; SPD, single-photon detector.}
	\label{fig:experimentalsetup}
\end{figure*}
%\del{\textit{Protocol—}A concise introduction to the three-user AMDI-QCC protocol is given {here, }with full details available in the Supplemental Materials. }
{The details of AMDI-QCC protocol, including state preparation, measurement, pairing, basis assignment, key mapping, parameter estimation, and finite-size key distillation, is provided in the Supplemental Material~\cite{SupplementalMaterial}.}
% 理论放在supplemental
%In each round, Alice prepares the coherent state pulse $\vert \sqrt{k_{A}}e^{\textbf{i}\phi_{A}}\rangle$ , where the intensity $k_A$ is randomly chosen from the set $\left\{\mu,\nu,0\right\}$, and the phase is randomly modulated from $\phi_{A}= \frac{2\pi}{16} \times n$. In this work, we set $0<\nu<\mu<1$ and $n = 0, 1, 2, \dots, 15$. Bob and Charlie similarly prepare coherent state pulses 
%$\vert \sqrt{k_{B}}e^{\textbf{i}\phi_{B}}\rangle$ and $\vert \sqrt{k_{C}}e^{\textbf{i}\phi_{C}}\rangle$
%, respectively. Subsequently, the three coherent state pulses are sent to the GHZ analyzer, Eve. Eve splits each incoming pulse from every user into two parts using a beam splitter (BS), and then performs interference measurements between the adjacent users’ pulses as shown in Fig.~\ref{fig:mainidea}(a). Eve publicly announces the outcomes of each round of interference measurement. After 
%$N$ rounds, the users retain only those detection events in which one and only one detector registers a click. Following the pairing strategy detailed in the Supplemental Materials, the basis choice and key value of each paired event are determined from the corresponding relative intensity and phase information. After performing basis sifting, key mapping, and data post-processing, the three-party AMDI-QCC protocol produces the final shared secret key, and 
Here, we provide the secure key length formula~\cite{lu2025repeater}:
\begin{equation}
	\begin{aligned}
		l \ge& \underline{s}_{111}^{Z}\left[1-H_2\left(\overline{e}_{111}^{Z,ph}\right)\right]\\
		&- f\overline{s}_{(\mu,\mu,\mu)}^{Z}\max\left[H_2\left(\overline{E}_{A,B}^{Z}\right),H_2\left(\overline{E}_{A,C}^{Z}\right)\right]\\
		& {
			- \log_2 (\frac{4}{\varepsilon_{\mathrm{cor}}})
			- 2 \log_2 (\frac{2}{\varepsilon' \hat{\varepsilon}})
			- 2 \log_2 \left(\frac{1}{2\varepsilon_{\mathrm{PA}}}\right)
		},
	\end{aligned}
\end{equation}
where $H_{2}(x)=-x\log_{2}x-(1-x)\log_{2}(1-x)$ denotes the binary entropy function, $f$ represents the error-correction efficiency. The single-photon contribution $s_{111}^{Z}$ of the $Z$-basis paired events, as well as the associated phase error rate $e_{111}^{Z,ph}$, can be estimated through the decoy-state method~\cite{lo2005decoy,wang2005beating}. The number of paired events used for final key distillation, $s^{Z}_{(\mu,\mu,\mu)}$, as well as the marginal bit error rates $E_{A,B}^{Z}$ and $E_{A,C}^{Z}$ with Alice’s key as the reference, can be directly obtained from the experimental data. Based on the Chernoff-Hoeffding method, $\underline{x}$ and $\overline{x}$ denote the lower and upper bounds of the observed value $x$, respectively. And $\varepsilon_{\mathrm{cor}}$ is the failure probability of error correction; $\varepsilon'$ and $\hat{\varepsilon}$ are the coefficients while using a chain rule for smooth entropies; $\varepsilon_{\mathrm{PA}}$ is the failure probability of privacy amplification. {Here, $\varepsilon_{\mathrm{cor}}=\varepsilon'=\hat{\varepsilon}=\varepsilon_{\mathrm{PA}}=\varepsilon=10^{-7}$}.

% end- 理论放supplemental

% 实验参数放进supplemental吧

%\begin{table}[b]
%	\centering
%	\caption{The table summarizes the total loss, channel attenuation, decoy-state parameters, error-correction efficiency, detector performance, and the corresponding secure key rates for total losses of 51.8~dB and 66.3~dB.
%		Here, the total loss is defined as the sum of the losses from all users to the central measurement node. The channel attenuation refers to the applied attenuation, excluding detection loss. The parameter $T_{}$ denotes the effective transmittance for a single user to the measurement node, including both channel loss and detection efficiency.
%	}
%	\label{tab:exp_parameters}
%	\begin{tabular}{lcc}
%		\hline\hline
%		Total loss (dB)		      & 51.80  & 66.3  \\
%		Total channel loss (dB)   & 35.48  & 51.24  \\
%		$\mu$                       & 0.2572 & 0.3535 \\
%		$\nu$                       & 0.0209 & 0.0413 \\
%		$p_\mu$                     & 0.15   & 0.15   \\
%		$p_\nu$                     & 0.35   & 0.35   \\
%		$f$ & 1.06 & 1.06 \\
%		Measurement Side loss (dB)     & 2.51   & 2.51   \\
%		Detector efficiency (\%)  & 81     & 81     \\
%		%Effective detector efficiency (\%) & 45 & 45 \\
%		% Indoor ratio              & 0.628  & 0.691  \\
%		% Effective efficiency with door (\%)& 28.25 & 31.05 \\
%		$T$                    & $1.88\times 10^{-2}$ & $6.18\times 10^{-3}$ \\
%		Key rate                  & $1.64\times 10^{-7}$ & $3.75\times 10^{-8}$ \\
%		\hline\hline
%	\end{tabular}
%\end{table}

% end-实验参数放supplemental

\begin{figure*}[http]
	\centering
	\includegraphics[width=0.8\linewidth]{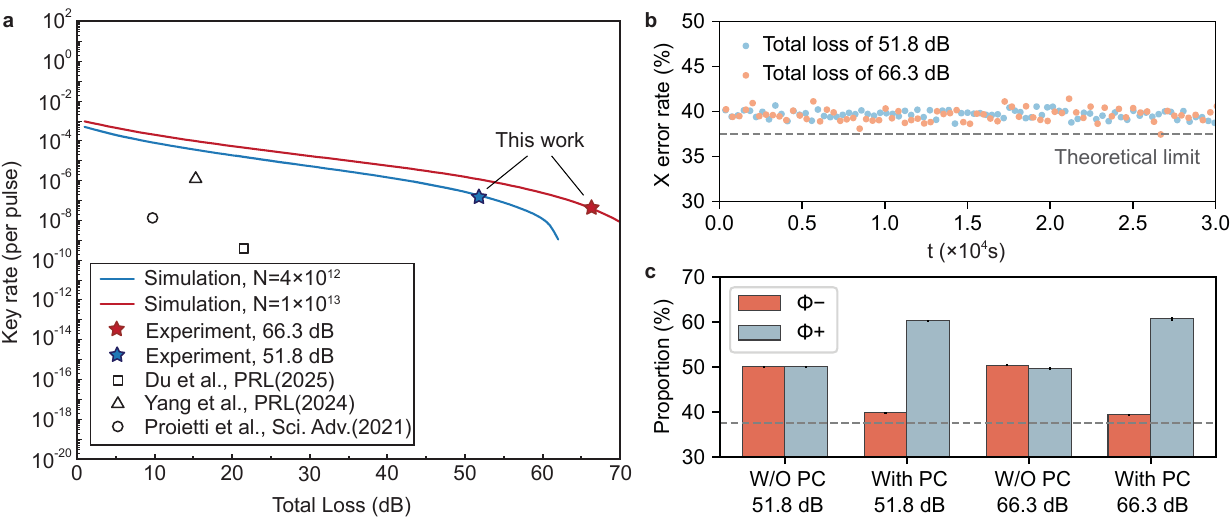} % 图片路径和文件名
	\caption{Key-rate performance and phase-compensation results.
		\textbf{a}. Secret key rate versus total channel loss, compared with previous experimental results~\cite{du2025experimental,yang2024experimental,proietti2021experimental}. 
		\textbf{b}. Stability of the X-basis error rate over time, with the theoretical lower bound (37.5\%) indicated. 
		\textbf{c}. Error-rate distribution with and without phase compensation at 51.8 dB and 66.3 dB total loss, where W/O PC denotes the case without phase compensation and With PC denotes the case with phase compensation.}
	\label{fig:keyrateresult}
\end{figure*}

The experimental setting is shown in Fig.~\ref{fig:experimentalsetup}, a single laser is split into three beams for Alice, Bob, and Charlie. For the case of independent lasers,~{which represent a crucial next step toward the practical implementation of AMDI-QCC,} the relative phases can be obtained by estimating the pairwise frequency differences between lasers\cite{zhu2023experimental,zhu2026frequency,shao2025high,zhang2025experimental}, with the detailed method provided in the Supplemental Material~\cite{SupplementalMaterial}. Each party modulates the continuous-wave light as follows: the light first passes through an intensity modulator to chop it into pulses with a repetition rate of 500\,MHz. These pulses then pass through a Sagnac loop for decoy-state modulation, where the pulses are randomly prepared in one of four intensity states: reference, signal, decoy, or vacuum. The reference pulses are used for phase compensation and are operated at a repetition rate of 288.49~MHz to enable accurate phase estimation. In the quantum-light zone, corresponding to the signal, decoy, and vacuum states, the pulses are operated at 143.52~MHz, where a phase modulator applies one of sixteen random phases, $\phi = \tfrac{2\pi}{16} \times n$ ($n = 0, 1, \ldots, 15$), to realize phase randomization. In the reference zone, certain pulses are modulated with additional phases of $\pi/2$ or $3\pi/2$, as detailed in the Supplemental Materials~\cite{SupplementalMaterial}. The modulation signals are generated by three arbitrary waveform generators (two Tektronix AWG70002B and one Keysight M8195A), synchronized via a shared clock distribution module and operated in trigger mode to ensure precise timing alignment among all channels. Finally, the pulses are attenuated by an electrically variable optical attenuator to the single-photon level before being sent to the detection site. {The detailed procedures for electrical signal synchronization and modulation, as well as optical modulation and the experimental parameters used in our implementation, are provided in the Supplemental Material~\cite{SupplementalMaterial}.}

At the detection site, photons from Alice, Bob, and Charlie are split and interfere pairwise at three detection ports. Polarization feedback is applied to all six optical paths by monitoring the non-interfering PBS outputs with single-photon detectors and adjusting the EPC voltages to minimize their counts, thereby aligning the photon polarizations. Temporal modes are aligned by statistically analyzing photon arrival times and tuning both the electronic delays of the arbitrary waveform generators and the optical delay lines, ensuring high-visibility interference. The insertion loss of the optical delay lines (0.24~dB) is included in the total channel loss. {Details of the feedback procedure and interference-visibility are provided in the Supplemental Material~\cite{SupplementalMaterial}.}

% put the Fig 4 to supplemental

%\begin{figure}[t]
%	\centering
%	\includegraphics[width=\linewidth]{fig_theta.pdf} % 或 .png/.jpg
%	\caption{
%		Single-photon interference and collective phase dependence among users.
%		(a–c) Error rate of single-photon interference as a function of the relative delay $\Delta t$ between interfering pulses for each user pair: 
%		(a) Alice–Bob, (b) Bob–Charlie, and (c) Charlie–Alice. \add{The error rate is evaluated as follows. For each detection event, the phase difference between two parties is calculated and classified as being near \(0\) or near \(\pi\). The expected detector is then determined accordingly, and an error is counted when the click occurs in the opposite detector.}
%		The error rate reaches a minimum at $\Delta t \approx 0$, indicating optimal temporal overlap and high-visibility interference. 
%		(d,e) Measured proportions of $\Phi^+$ and $\Phi^-$ components as functions of the relative phase-difference signs $(\theta_B - \theta_A)$, $(\theta_C - \theta_B)$, and $(\theta_A - \theta_C)$. 
%		Each cube vertex represents one combination of sign conventions (“$+$” or “$-$”) corresponding to the measured proportion. 
%		“No flip” denotes the physically correct choice of phase-difference sign, while “flip” indicates an incorrect sign assignment. 
%		The results highlight the collective phase dependence among the three parties in GHZ-state measurement.
%	}
%	\label{fig:cube-phi}
%\end{figure}

We collect the detection events from all six single-photon detectors and apply a sliding-window pairing algorithm that processes the data port-by-port (ports 1, 2, and 3) rather than in chronological order. The port-by-port strategy eliminates the need to distinguish which detector clicks first, simplifying the subsequent analysis, while the sliding-window algorithm improves the overall pairing efficiency, as detailed in Supplemental Materials~\cite{SupplementalMaterial}.

%\del{At both 51.8~dB and 66.3~dB total loss, the maximum pairing length was set to 100~$\mu$s.  A detailed discussion on the choice of maximum pairing length is provided in the Supplemental Material. With a 100~$\mu$s acquisition window and a system clock rate of 500~MHz, most of the detection clicks can be paired for both total-loss conditions, given the corresponding transmittances~$T$. The obtained pairs are assigned to the $X$- and $Z$- basis according to the protocol rules, where the $Z$-basis is used for secure key generation and the $X$-basis is employed for phase-error estimation and eavesdropping detection. Further implementation details are provided in the Supplemental Materials.}

Notably, reliable GHZ-state measurements require accurate estimation of the phase drift of the six optical paths. To this end, we use the reference pulses to estimate and compensate the fiber phase, and incorporate the extracted phases into parameter estimation, particularly for the $X$-basis error rate. Importantly, the {signs of the three phase differences among the users (e.g., $\theta_B-\theta_A$, $\theta_C-\theta_B$, and $\theta_A-\theta_C$) must be pre-calibrated. {We further present the GHZ-state error rates under different sign conventions in the Supplemental Materials~\cite{SupplementalMaterial}. The GHZ-state measurement error is minimized only when the phase of each user is correctly estimated, highlighting the collective phase dependence among the three parties. {The details for determining the correct signs of these phase differences are provided in the Supplemental Material.~\cite{SupplementalMaterial}}

The experiments were carried out under total losses of 51.8~dB and 66.3~dB. At 51.8~dB, the $Z$- and $X$-basis error rates were $0.20\%$ and $39.68\%$, respectively; at 66.3~dB, they were $0.10\%$ and $39.42\%$. {Given the quantum-light repetition rate of 143.52~MHz, the secure key generation rates are 23.5~bit/s and 5.39~bit/s at total losses of 51.8~dB and 66.3~dB, corresponding to per-pulse secure key rates of \(R_1=1.64\times10^{-7}\) and \(R_2=3.75\times10^{-8}\), respectively. The corresponding single-arm channel losses are 17.2~dB and 22.1~dB, giving single-arm channel transmittances of \(T_1=1.88\times10^{-2}\) and \(T_2=6.18\times10^{-3}\). We therefore obtain \(R_1/T_1=8.59\times10^{-6}\) and \(R_2/T_2=6.08\times10^{-6}\). The ratio of $R/T$, together with Fig.~\ref{fig:keyrateresult}(a), provides supporting evidence for the expected near-linear $O(T)$ scaling,} \del{}representing a significant improvement over the conventional multi-user scaling of $O(T^{N})$~\cite{fu2015long,du2025experimental,yang2024experimental}.

%Notably, the obtained key rates surpass the repeaterless multi-end communication bound~~\cite{pirandola2019bounds,pirandola2019end}, demonstrating the advantage of our implementation in efficiently utilizing single-photon interference even under high channel losses. 

%\add{Figure~\ref{fig:keyrateresult}(a) shows that the measured key rate follows the expected approximately linear dependence on the transmittance \(T\), providing supporting evidence for the predicted scaling behavior of the AMDI-QCC protocol.}\del{Figure~\ref{fig:keyrateresult}(a) shows that the key rate scales approximately linearly with the transmittance~$T$, confirming the expected performance of the AMDI-QCC protocol.} 

In Fig.~\ref{fig:keyrateresult}(b), the $X$-basis error rate fluctuates around the theoretical lower bound of $37.5\%$~\cite{fu2015long}. Owing to the presence of multi-photon components, this value represents the fundamental limit of the $X$-basis error. The experimental discrepancy is mainly attributed to residual fiber phase-estimation errors, imperfect polarization alignment, temporal and spectral mismatch, and phase-modulation misalignment. The long-term stability of the fluctuation further indicates that our system maintains stable phase coherence over extended operation periods. Due to our time-division-multiplexed phase compensation, the measured error approaches this bound.  
As shown in Fig.~\ref{fig:keyrateresult}(c), the GHZ-state error rate strongly depends on the precision of phase compensation among all users. When the total phase is stabilized to $0$ or $\pi$, the system deterministically projects onto $|\Phi^+\rangle$ or $|\Phi^-\rangle$, producing two distinct interference outcomes that correspond to complementary detector-response distributions. Any residual phase deviation blurs this distinction, reducing interference visibility and increasing X-basis error. Therefore, maximizing the phase contrast between these two GHZ-state projections is crucial for achieving high-fidelity multi-party interference and stable key generation.

In conclusion, we have carried out QCC at the highest channel loss achieved by implementing simultaneous three-party phase, time, and polarization alignment to enable high interference visibility. By designing an effective pairing strategy and carefully analyzing the relation between the three-party phase and GHZ-state visibility, we minimized the GHZ-state measurement error. With these techniques, we successfully demonstrated the AMDI-QCC experiment. Especially, at total channel losses of 51.8~dB and 66.3~dB, we obtained secret key rates of $1.64\times10^{-7}$ and $3.75\times10^{-8}$ per pulse, corresponding to 23.5 and 5.4~bit/s. {These results provide supporting evidence that the key-rate decay at two channel losses follows an approximately \(O(T)\) scaling~\cite{lu2025repeater}.}

For the case of $N>3$, the extension of our scheme is straightforward. Each additional user only requires one extra interference module at the measurement station. Correspondingly, phase stabilization introduces only a linear overhead, involving one additional path-phase feedback and one extra frequency-difference. A detailed analysis of this extension is provided in the Supplemental Material. {Compared with pairwise QKD followed by bitwise XOR{~\cite{lu2025fully}}, the AMDI-QCC protocol establishes the conference key through a single multi-user interference process, avoiding the need for up to $N-1$ independent pairwise QKD sessions and reducing the communication-resource overhead in larger networks.}

Our approach demonstrates the feasibility of long-distance, phase-stable, multi-user QCC based on single-photon interference and the efficient pairing strategy, without the need for entanglement distribution or quantum repeaters. The methods developed here—including multi-party phase compensation and optimized detection pairing—can be directly extended to larger-scale user networks and integrated photonic platforms. These results establish a practical benchmark for the future realization of {scalable} quantum communication networks. 

Looking forward,~{implementing AMDI-QCC with independent lasers is an important next step. Besides,} in practical quantum networks, the distances from different users to the detection site are typically unequal, indicating that future studies could explore the performance of our scheme under asymmetric channel conditions. Furthermore, our experimental approach can be readily extended to include more users, substantially reducing resource consumption in multi-user quantum networks. Beyond multi-party quantum communication, efficient GHZ-state measurement can also benefit a wide range of quantum technologies, including multi-user network synchronization, quantum sensing, quantum network coherence, and measurement-based quantum computing. 

%\textit{Acknowledgements—}
%This work was supported by A*STAR (M21K2c0116, M24M8b0004), the Singapore National Research Foundation (NRF-CRP22-2019-0004, NRF-CRP30-2023-0003, NRF-CRP31-0001, NRF2023-ITC004-001, and NRF-MSG-2023-0002), and the Singapore Ministry of Education Tier 2 Grant (MOE-T2EP50221-0005, MOE-T2EP50222-0018). We also acknowledge the support from the Zhejiang Provincial Natural Science Foundation of China Grant (No.~LQ24A050005) and the Quantum Science and Technology - National Science and Technology Major Project (Grant No.~2024ZD0302200)

%\begin{acknowledgments}
	
\textit{Acknowledgments—}This work was supported by A*STAR (M21K2c0116, M24M8b0004), the Singapore National Research Foundation (NRF-CRP22-2019-0004, NRF-CRP30-2023-0003, NRF-CRP31-0001, NRF2023-ITC004-001, and NRF-MSG-2023-0002), and the Singapore Ministry of Education Tier 2 Grant (MOE-T2EP50221-0005, MOE-T2EP50222-0018). We also acknowledge the support from the Zhejiang Provincial Natural Science Foundation of China Grant (No.~LQ24A050005) and the Quantum Science and Technology - National Science and Technology Major Project (Grant No.~2024ZD0302200)
\nocite{fang2020implementation,liu2023experimental,wang2022twin,chen2020sending,chen2024twin,zhang2017improved}

\bibliography{apssamp}% Produces the bibliography via BibTeX.

\begin{thebibliography}{40}%
\makeatletter
\providecommand \@ifxundefined [1]{%
 \@ifx{#1\undefined}
}%
\providecommand \@ifnum [1]{%
 \ifnum #1\expandafter \@firstoftwo
 \else \expandafter \@secondoftwo
 \fi
}%
\providecommand \@ifx [1]{%
 \ifx #1\expandafter \@firstoftwo
 \else \expandafter \@secondoftwo
 \fi
}%
\providecommand \natexlab [1]{#1}%
\providecommand \enquote  [1]{``#1''}%
\providecommand \bibnamefont  [1]{#1}%
\providecommand \bibfnamefont [1]{#1}%
\providecommand \citenamefont [1]{#1}%
\providecommand \href@noop [0]{\@secondoftwo}%
\providecommand \href [0]{\begingroup \@sanitize@url \@href}%
\providecommand \@href[1]{\@@startlink{#1}\@@href}%
\providecommand \@@href[1]{\endgroup#1\@@endlink}%
\providecommand \@sanitize@url [0]{\catcode `\\12\catcode `\$12\catcode
  `\&12\catcode `\#12\catcode `\^12\catcode `\_12\catcode `\%12\relax}%
\providecommand \@@startlink[1]{}%
\providecommand \@@endlink[0]{}%
\providecommand \url  [0]{\begingroup\@sanitize@url \@url }%
\providecommand \@url [1]{\endgroup\@href {#1}{\urlprefix }}%
\providecommand \urlprefix  [0]{URL }%
\providecommand \Eprint [0]{\href }%
\providecommand \doibase [0]{https://doi.org/}%
\providecommand \selectlanguage [0]{\@gobble}%
\providecommand \bibinfo  [0]{\@secondoftwo}%
\providecommand \bibfield  [0]{\@secondoftwo}%
\providecommand \translation [1]{[#1]}%
\providecommand \BibitemOpen [0]{}%
\providecommand \bibitemStop [0]{}%
\providecommand \bibitemNoStop [0]{.\EOS\space}%
\providecommand \EOS [0]{\spacefactor3000\relax}%
\providecommand \BibitemShut  [1]{\csname bibitem#1\endcsname}%
\let\auto@bib@innerbib\@empty
%</preamble>
\bibitem [{\citenamefont {Kimble}(2008)}]{kimble2008quantum}%
  \BibitemOpen
  \bibfield  {author} {\bibinfo {author} {\bibfnamefont {H.~J.}\ \bibnamefont
  {Kimble}},\ }\bibfield  {title} {\bibinfo {title} {The quantum internet},\
  }\href@noop {} {\bibfield  {journal} {\bibinfo  {journal} {Nature}\ }\textbf
  {\bibinfo {volume} {453}},\ \bibinfo {pages} {1023} (\bibinfo {year}
  {2008})}\BibitemShut {NoStop}%
\bibitem [{\citenamefont {Wehner}\ \emph {et~al.}(2018)\citenamefont {Wehner},
  \citenamefont {Elkouss},\ and\ \citenamefont {Hanson}}]{wehner2018quantum}%
  \BibitemOpen
  \bibfield  {author} {\bibinfo {author} {\bibfnamefont {S.}~\bibnamefont
  {Wehner}}, \bibinfo {author} {\bibfnamefont {D.}~\bibnamefont {Elkouss}},\
  and\ \bibinfo {author} {\bibfnamefont {R.}~\bibnamefont {Hanson}},\
  }\bibfield  {title} {\bibinfo {title} {Quantum internet: A vision for the
  road ahead},\ }\href@noop {} {\bibfield  {journal} {\bibinfo  {journal}
  {Science}\ }\textbf {\bibinfo {volume} {362}},\ \bibinfo {pages} {eaam9288}
  (\bibinfo {year} {2018})}\BibitemShut {NoStop}%
\bibitem [{\citenamefont {Pittaluga}\ \emph {et~al.}(2025)\citenamefont
  {Pittaluga}, \citenamefont {Lo}, \citenamefont {Brzosko}, \citenamefont
  {Woodward}, \citenamefont {Scalcon}, \citenamefont {Winnel}, \citenamefont
  {Roger}, \citenamefont {Dynes}, \citenamefont {Owen}, \citenamefont
  {Ju{\'a}rez} \emph {et~al.}}]{pittaluga2025long}%
  \BibitemOpen
  \bibfield  {author} {\bibinfo {author} {\bibfnamefont {M.}~\bibnamefont
  {Pittaluga}}, \bibinfo {author} {\bibfnamefont {Y.~S.}\ \bibnamefont {Lo}},
  \bibinfo {author} {\bibfnamefont {A.}~\bibnamefont {Brzosko}}, \bibinfo
  {author} {\bibfnamefont {R.~I.}\ \bibnamefont {Woodward}}, \bibinfo {author}
  {\bibfnamefont {D.}~\bibnamefont {Scalcon}}, \bibinfo {author} {\bibfnamefont
  {M.~S.}\ \bibnamefont {Winnel}}, \bibinfo {author} {\bibfnamefont
  {T.}~\bibnamefont {Roger}}, \bibinfo {author} {\bibfnamefont {J.~F.}\
  \bibnamefont {Dynes}}, \bibinfo {author} {\bibfnamefont {K.~A.}\ \bibnamefont
  {Owen}}, \bibinfo {author} {\bibfnamefont {S.}~\bibnamefont {Ju{\'a}rez}},
  \emph {et~al.},\ }\bibfield  {title} {\bibinfo {title} {Long-distance
  coherent quantum communications in deployed telecom networks},\ }\href@noop
  {} {\bibfield  {journal} {\bibinfo  {journal} {Nature}\ }\textbf {\bibinfo
  {volume} {640}},\ \bibinfo {pages} {911} (\bibinfo {year}
  {2025})}\BibitemShut {NoStop}%
\bibitem [{\citenamefont {Bennett}\ and\ \citenamefont
  {Brassard}(2014)}]{bennett2014quantum}%
  \BibitemOpen
  \bibfield  {author} {\bibinfo {author} {\bibfnamefont {C.~H.}\ \bibnamefont
  {Bennett}}\ and\ \bibinfo {author} {\bibfnamefont {G.}~\bibnamefont
  {Brassard}},\ }\bibfield  {title} {\bibinfo {title} {Quantum cryptography:
  Public key distribution and coin tossing},\ }\href@noop {} {\bibfield
  {journal} {\bibinfo  {journal} {Theoretical Computer Science}\ }\textbf
  {\bibinfo {volume} {560}},\ \bibinfo {pages} {7} (\bibinfo {year}
  {2014})}\BibitemShut {NoStop}%
\bibitem [{\citenamefont {Ekert}(1991)}]{ekert1991quantum}%
  \BibitemOpen
  \bibfield  {author} {\bibinfo {author} {\bibfnamefont {A.~K.}\ \bibnamefont
  {Ekert}},\ }\bibfield  {title} {\bibinfo {title} {Quantum cryptography based
  on bell’s theorem},\ }\href@noop {} {\bibfield  {journal} {\bibinfo
  {journal} {Physical Review Letters}\ }\textbf {\bibinfo {volume} {67}},\
  \bibinfo {pages} {661} (\bibinfo {year} {1991})}\BibitemShut {NoStop}%
\bibitem [{\citenamefont {Gisin}\ \emph {et~al.}(2002)\citenamefont {Gisin},
  \citenamefont {Ribordy}, \citenamefont {Tittel},\ and\ \citenamefont
  {Zbinden}}]{gisin2002quantum}%
  \BibitemOpen
  \bibfield  {author} {\bibinfo {author} {\bibfnamefont {N.}~\bibnamefont
  {Gisin}}, \bibinfo {author} {\bibfnamefont {G.}~\bibnamefont {Ribordy}},
  \bibinfo {author} {\bibfnamefont {W.}~\bibnamefont {Tittel}},\ and\ \bibinfo
  {author} {\bibfnamefont {H.}~\bibnamefont {Zbinden}},\ }\bibfield  {title}
  {\bibinfo {title} {Quantum cryptography},\ }\href@noop {} {\bibfield
  {journal} {\bibinfo  {journal} {Reviews of Modern Physics}\ }\textbf
  {\bibinfo {volume} {74}},\ \bibinfo {pages} {145} (\bibinfo {year}
  {2002})}\BibitemShut {NoStop}%
\bibitem [{\citenamefont {Scarani}\ \emph {et~al.}(2009)\citenamefont
  {Scarani}, \citenamefont {Bechmann-Pasquinucci}, \citenamefont {Cerf},
  \citenamefont {Du{\v{s}}ek}, \citenamefont {L{\"u}tkenhaus},\ and\
  \citenamefont {Peev}}]{scarani2009security}%
  \BibitemOpen
  \bibfield  {author} {\bibinfo {author} {\bibfnamefont {V.}~\bibnamefont
  {Scarani}}, \bibinfo {author} {\bibfnamefont {H.}~\bibnamefont
  {Bechmann-Pasquinucci}}, \bibinfo {author} {\bibfnamefont {N.~J.}\
  \bibnamefont {Cerf}}, \bibinfo {author} {\bibfnamefont {M.}~\bibnamefont
  {Du{\v{s}}ek}}, \bibinfo {author} {\bibfnamefont {N.}~\bibnamefont
  {L{\"u}tkenhaus}},\ and\ \bibinfo {author} {\bibfnamefont {M.}~\bibnamefont
  {Peev}},\ }\bibfield  {title} {\bibinfo {title} {The security of practical
  quantum key distribution},\ }\href@noop {} {\bibfield  {journal} {\bibinfo
  {journal} {Reviews of Modern Physics}\ }\textbf {\bibinfo {volume} {81}},\
  \bibinfo {pages} {1301} (\bibinfo {year} {2009})}\BibitemShut {NoStop}%
\bibitem [{\citenamefont {Xu}\ \emph {et~al.}(2020)\citenamefont {Xu},
  \citenamefont {Ma}, \citenamefont {Zhang}, \citenamefont {Lo},\ and\
  \citenamefont {Pan}}]{xu2020secure}%
  \BibitemOpen
  \bibfield  {author} {\bibinfo {author} {\bibfnamefont {F.}~\bibnamefont
  {Xu}}, \bibinfo {author} {\bibfnamefont {X.}~\bibnamefont {Ma}}, \bibinfo
  {author} {\bibfnamefont {Q.}~\bibnamefont {Zhang}}, \bibinfo {author}
  {\bibfnamefont {H.-K.}\ \bibnamefont {Lo}},\ and\ \bibinfo {author}
  {\bibfnamefont {J.-W.}\ \bibnamefont {Pan}},\ }\bibfield  {title} {\bibinfo
  {title} {Secure quantum key distribution with realistic devices},\
  }\href@noop {} {\bibfield  {journal} {\bibinfo  {journal} {Reviews of Modern
  Physics}\ }\textbf {\bibinfo {volume} {92}},\ \bibinfo {pages} {025002}
  (\bibinfo {year} {2020})}\BibitemShut {NoStop}%
\bibitem [{\citenamefont {Epping}\ \emph {et~al.}(2017)\citenamefont {Epping},
  \citenamefont {Kampermann}, \citenamefont {Bru{\ss}} \emph
  {et~al.}}]{epping2017multi}%
  \BibitemOpen
  \bibfield  {author} {\bibinfo {author} {\bibfnamefont {M.}~\bibnamefont
  {Epping}}, \bibinfo {author} {\bibfnamefont {H.}~\bibnamefont {Kampermann}},
  \bibinfo {author} {\bibfnamefont {D.}~\bibnamefont {Bru{\ss}}}, \emph
  {et~al.},\ }\bibfield  {title} {\bibinfo {title} {Multi-partite entanglement
  can speed up quantum key distribution in networks},\ }\href@noop {}
  {\bibfield  {journal} {\bibinfo  {journal} {New Journal of Physics}\ }\textbf
  {\bibinfo {volume} {19}},\ \bibinfo {pages} {093012} (\bibinfo {year}
  {2017})}\BibitemShut {NoStop}%
\bibitem [{\citenamefont {Bose}\ \emph {et~al.}(1998)\citenamefont {Bose},
  \citenamefont {Vedral},\ and\ \citenamefont
  {Knight}}]{bose1998multiparticle}%
  \BibitemOpen
  \bibfield  {author} {\bibinfo {author} {\bibfnamefont {S.}~\bibnamefont
  {Bose}}, \bibinfo {author} {\bibfnamefont {V.}~\bibnamefont {Vedral}},\ and\
  \bibinfo {author} {\bibfnamefont {P.~L.}\ \bibnamefont {Knight}},\ }\bibfield
   {title} {\bibinfo {title} {Multiparticle generalization of entanglement
  swapping},\ }\href@noop {} {\bibfield  {journal} {\bibinfo  {journal}
  {Physical Review A}\ }\textbf {\bibinfo {volume} {57}},\ \bibinfo {pages}
  {822} (\bibinfo {year} {1998})}\BibitemShut {NoStop}%
\bibitem [{\citenamefont {Chen}\ and\ \citenamefont
  {Lo}(2004)}]{chen2004multi}%
  \BibitemOpen
  \bibfield  {author} {\bibinfo {author} {\bibfnamefont {K.}~\bibnamefont
  {Chen}}\ and\ \bibinfo {author} {\bibfnamefont {H.-K.}\ \bibnamefont {Lo}},\
  }\bibfield  {title} {\bibinfo {title} {Multi-partite quantum cryptographic
  protocols with noisy ghz states},\ }\href@noop {} {\bibfield  {journal}
  {\bibinfo  {journal} {arXiv preprint quant-ph/0404133}\ } (\bibinfo {year}
  {2004})}\BibitemShut {NoStop}%
\bibitem [{\citenamefont {Murta}\ \emph {et~al.}(2020)\citenamefont {Murta},
  \citenamefont {Grasselli}, \citenamefont {Kampermann},\ and\ \citenamefont
  {Bru{\ss}}}]{murta2020quantum}%
  \BibitemOpen
  \bibfield  {author} {\bibinfo {author} {\bibfnamefont {G.}~\bibnamefont
  {Murta}}, \bibinfo {author} {\bibfnamefont {F.}~\bibnamefont {Grasselli}},
  \bibinfo {author} {\bibfnamefont {H.}~\bibnamefont {Kampermann}},\ and\
  \bibinfo {author} {\bibfnamefont {D.}~\bibnamefont {Bru{\ss}}},\ }\bibfield
  {title} {\bibinfo {title} {Quantum conference key agreement: A review},\
  }\href@noop {} {\bibfield  {journal} {\bibinfo  {journal} {Advanced Quantum
  Technologies}\ }\textbf {\bibinfo {volume} {3}},\ \bibinfo {pages} {2000025}
  (\bibinfo {year} {2020})}\BibitemShut {NoStop}%
\bibitem [{\citenamefont {Grasselli}\ \emph {et~al.}(2018)\citenamefont
  {Grasselli}, \citenamefont {Kampermann},\ and\ \citenamefont
  {Bru{\ss}}}]{grasselli2018finite}%
  \BibitemOpen
  \bibfield  {author} {\bibinfo {author} {\bibfnamefont {F.}~\bibnamefont
  {Grasselli}}, \bibinfo {author} {\bibfnamefont {H.}~\bibnamefont
  {Kampermann}},\ and\ \bibinfo {author} {\bibfnamefont {D.}~\bibnamefont
  {Bru{\ss}}},\ }\bibfield  {title} {\bibinfo {title} {Finite-key effects in
  multipartite quantum key distribution protocols},\ }\href@noop {} {\bibfield
  {journal} {\bibinfo  {journal} {New Journal of Physics}\ }\textbf {\bibinfo
  {volume} {20}},\ \bibinfo {pages} {113014} (\bibinfo {year}
  {2018})}\BibitemShut {NoStop}%
\bibitem [{\citenamefont {Fu}\ \emph {et~al.}(2015)\citenamefont {Fu},
  \citenamefont {Yin}, \citenamefont {Chen},\ and\ \citenamefont
  {Chen}}]{fu2015long}%
  \BibitemOpen
  \bibfield  {author} {\bibinfo {author} {\bibfnamefont {Y.}~\bibnamefont
  {Fu}}, \bibinfo {author} {\bibfnamefont {H.-L.}\ \bibnamefont {Yin}},
  \bibinfo {author} {\bibfnamefont {T.-Y.}\ \bibnamefont {Chen}},\ and\
  \bibinfo {author} {\bibfnamefont {Z.-B.}\ \bibnamefont {Chen}},\ }\bibfield
  {title} {\bibinfo {title} {Long-distance measurement-device-independent
  multiparty quantum communication},\ }\href@noop {} {\bibfield  {journal}
  {\bibinfo  {journal} {Physical Review Letters}\ }\textbf {\bibinfo {volume}
  {114}},\ \bibinfo {pages} {090501} (\bibinfo {year} {2015})}\BibitemShut
  {NoStop}%
\bibitem [{\citenamefont {Zhao}\ \emph {et~al.}(2020)\citenamefont {Zhao},
  \citenamefont {Zeng}, \citenamefont {Cao}, \citenamefont {Xu}, \citenamefont
  {Zhen}, \citenamefont {Ma}, \citenamefont {Li}, \citenamefont {Liu},\ and\
  \citenamefont {Chen}}]{zhao2020phase}%
  \BibitemOpen
  \bibfield  {author} {\bibinfo {author} {\bibfnamefont {S.}~\bibnamefont
  {Zhao}}, \bibinfo {author} {\bibfnamefont {P.}~\bibnamefont {Zeng}}, \bibinfo
  {author} {\bibfnamefont {W.-F.}\ \bibnamefont {Cao}}, \bibinfo {author}
  {\bibfnamefont {X.-Y.}\ \bibnamefont {Xu}}, \bibinfo {author} {\bibfnamefont
  {Y.-Z.}\ \bibnamefont {Zhen}}, \bibinfo {author} {\bibfnamefont
  {X.}~\bibnamefont {Ma}}, \bibinfo {author} {\bibfnamefont {L.}~\bibnamefont
  {Li}}, \bibinfo {author} {\bibfnamefont {N.-L.}\ \bibnamefont {Liu}},\ and\
  \bibinfo {author} {\bibfnamefont {K.}~\bibnamefont {Chen}},\ }\bibfield
  {title} {\bibinfo {title} {Phase-matching quantum cryptographic
  conferencing},\ }\href@noop {} {\bibfield  {journal} {\bibinfo  {journal}
  {Physical Review Applied}\ }\textbf {\bibinfo {volume} {14}},\ \bibinfo
  {pages} {024010} (\bibinfo {year} {2020})}\BibitemShut {NoStop}%
\bibitem [{\citenamefont {Cao}\ \emph {et~al.}(2021)\citenamefont {Cao},
  \citenamefont {Gu}, \citenamefont {Lu}, \citenamefont {Yin},\ and\
  \citenamefont {Chen}}]{cao2021coherent}%
  \BibitemOpen
  \bibfield  {author} {\bibinfo {author} {\bibfnamefont {X.-Y.}\ \bibnamefont
  {Cao}}, \bibinfo {author} {\bibfnamefont {J.}~\bibnamefont {Gu}}, \bibinfo
  {author} {\bibfnamefont {Y.-S.}\ \bibnamefont {Lu}}, \bibinfo {author}
  {\bibfnamefont {H.-L.}\ \bibnamefont {Yin}},\ and\ \bibinfo {author}
  {\bibfnamefont {Z.-B.}\ \bibnamefont {Chen}},\ }\bibfield  {title} {\bibinfo
  {title} {Coherent one-way quantum conference key agreement based on twin
  field},\ }\href@noop {} {\bibfield  {journal} {\bibinfo  {journal} {New
  Journal of Physics}\ }\textbf {\bibinfo {volume} {23}},\ \bibinfo {pages}
  {043002} (\bibinfo {year} {2021})}\BibitemShut {NoStop}%
\bibitem [{\citenamefont {Carrara}\ \emph {et~al.}(2023)\citenamefont
  {Carrara}, \citenamefont {Murta},\ and\ \citenamefont
  {Grasselli}}]{carrara2023overcoming}%
  \BibitemOpen
  \bibfield  {author} {\bibinfo {author} {\bibfnamefont {G.}~\bibnamefont
  {Carrara}}, \bibinfo {author} {\bibfnamefont {G.}~\bibnamefont {Murta}},\
  and\ \bibinfo {author} {\bibfnamefont {F.}~\bibnamefont {Grasselli}},\
  }\bibfield  {title} {\bibinfo {title} {Overcoming fundamental bounds on
  quantum conference key agreement},\ }\href@noop {} {\bibfield  {journal}
  {\bibinfo  {journal} {Physical Review Applied}\ }\textbf {\bibinfo {volume}
  {19}},\ \bibinfo {pages} {064017} (\bibinfo {year} {2023})}\BibitemShut
  {NoStop}%
\bibitem [{\citenamefont {Proietti}\ \emph {et~al.}(2021)\citenamefont
  {Proietti}, \citenamefont {Ho}, \citenamefont {Grasselli}, \citenamefont
  {Barrow}, \citenamefont {Malik},\ and\ \citenamefont
  {Fedrizzi}}]{proietti2021experimental}%
  \BibitemOpen
  \bibfield  {author} {\bibinfo {author} {\bibfnamefont {M.}~\bibnamefont
  {Proietti}}, \bibinfo {author} {\bibfnamefont {J.}~\bibnamefont {Ho}},
  \bibinfo {author} {\bibfnamefont {F.}~\bibnamefont {Grasselli}}, \bibinfo
  {author} {\bibfnamefont {P.}~\bibnamefont {Barrow}}, \bibinfo {author}
  {\bibfnamefont {M.}~\bibnamefont {Malik}},\ and\ \bibinfo {author}
  {\bibfnamefont {A.}~\bibnamefont {Fedrizzi}},\ }\bibfield  {title} {\bibinfo
  {title} {Experimental quantum conference key agreement},\ }\href@noop {}
  {\bibfield  {journal} {\bibinfo  {journal} {Science Advances}\ }\textbf
  {\bibinfo {volume} {7}},\ \bibinfo {pages} {eabe0395} (\bibinfo {year}
  {2021})}\BibitemShut {NoStop}%
\bibitem [{\citenamefont {Hahn}\ \emph {et~al.}(2020)\citenamefont {Hahn},
  \citenamefont {de~Jong},\ and\ \citenamefont {Pappa}}]{hahn2020anonymous}%
  \BibitemOpen
  \bibfield  {author} {\bibinfo {author} {\bibfnamefont {F.}~\bibnamefont
  {Hahn}}, \bibinfo {author} {\bibfnamefont {J.}~\bibnamefont {de~Jong}},\ and\
  \bibinfo {author} {\bibfnamefont {A.}~\bibnamefont {Pappa}},\ }\bibfield
  {title} {\bibinfo {title} {Anonymous quantum conference key agreement},\
  }\href@noop {} {\bibfield  {journal} {\bibinfo  {journal} {PRX Quantum}\
  }\textbf {\bibinfo {volume} {1}},\ \bibinfo {pages} {020325} (\bibinfo {year}
  {2020})}\BibitemShut {NoStop}%
\bibitem [{\citenamefont {Grasselli}\ \emph {et~al.}(2019)\citenamefont
  {Grasselli}, \citenamefont {Kampermann},\ and\ \citenamefont
  {Bru{\ss}}}]{grasselli2019conference}%
  \BibitemOpen
  \bibfield  {author} {\bibinfo {author} {\bibfnamefont {F.}~\bibnamefont
  {Grasselli}}, \bibinfo {author} {\bibfnamefont {H.}~\bibnamefont
  {Kampermann}},\ and\ \bibinfo {author} {\bibfnamefont {D.}~\bibnamefont
  {Bru{\ss}}},\ }\bibfield  {title} {\bibinfo {title} {Conference key agreement
  with single-photon interference},\ }\href@noop {} {\bibfield  {journal}
  {\bibinfo  {journal} {New Journal of Physics}\ }\textbf {\bibinfo {volume}
  {21}},\ \bibinfo {pages} {123002} (\bibinfo {year} {2019})}\BibitemShut
  {NoStop}%
\bibitem [{\citenamefont {Lu}\ \emph {et~al.}(2025{\natexlab{a}})\citenamefont
  {Lu}, \citenamefont {Yin}, \citenamefont {Xie}, \citenamefont {Fu},\ and\
  \citenamefont {Chen}}]{lu2025repeater}%
  \BibitemOpen
  \bibfield  {author} {\bibinfo {author} {\bibfnamefont {Y.-S.}\ \bibnamefont
  {Lu}}, \bibinfo {author} {\bibfnamefont {H.-L.}\ \bibnamefont {Yin}},
  \bibinfo {author} {\bibfnamefont {Y.-M.}\ \bibnamefont {Xie}}, \bibinfo
  {author} {\bibfnamefont {Y.}~\bibnamefont {Fu}},\ and\ \bibinfo {author}
  {\bibfnamefont {Z.-B.}\ \bibnamefont {Chen}},\ }\bibfield  {title} {\bibinfo
  {title} {Repeater-like asynchronous measurement-device-independent quantum
  conference key agreement},\ }\href@noop {} {\bibfield  {journal} {\bibinfo
  {journal} {Reports on Progress in Physics}\ }\textbf {\bibinfo {volume}
  {88}},\ \bibinfo {pages} {067901} (\bibinfo {year}
  {2025}{\natexlab{a}})}\BibitemShut {NoStop}%
\bibitem [{\citenamefont {Xie}\ \emph {et~al.}(2024)\citenamefont {Xie},
  \citenamefont {Lu}, \citenamefont {Fu}, \citenamefont {Yin},\ and\
  \citenamefont {Chen}}]{xie2024multi}%
  \BibitemOpen
  \bibfield  {author} {\bibinfo {author} {\bibfnamefont {Y.-M.}\ \bibnamefont
  {Xie}}, \bibinfo {author} {\bibfnamefont {Y.-S.}\ \bibnamefont {Lu}},
  \bibinfo {author} {\bibfnamefont {Y.}~\bibnamefont {Fu}}, \bibinfo {author}
  {\bibfnamefont {H.-L.}\ \bibnamefont {Yin}},\ and\ \bibinfo {author}
  {\bibfnamefont {Z.-B.}\ \bibnamefont {Chen}},\ }\bibfield  {title} {\bibinfo
  {title} {Multi-field quantum conferencing overcomes the network capacity
  limit},\ }\href@noop {} {\bibfield  {journal} {\bibinfo  {journal}
  {Communications Physics}\ }\textbf {\bibinfo {volume} {7}},\ \bibinfo {pages}
  {410} (\bibinfo {year} {2024})}\BibitemShut {NoStop}%
\bibitem [{\citenamefont {Greenberger}\ \emph {et~al.}(1989)\citenamefont
  {Greenberger}, \citenamefont {Horne},\ and\ \citenamefont
  {Zeilinger}}]{greenberger1989going}%
  \BibitemOpen
  \bibfield  {author} {\bibinfo {author} {\bibfnamefont {D.~M.}\ \bibnamefont
  {Greenberger}}, \bibinfo {author} {\bibfnamefont {M.~A.}\ \bibnamefont
  {Horne}},\ and\ \bibinfo {author} {\bibfnamefont {A.}~\bibnamefont
  {Zeilinger}},\ }\bibfield  {title} {\bibinfo {title} {Going beyond bell’s
  theorem},\ }\bibfield  {booktitle} {\emph {\bibinfo {booktitle} {Bell’s
  theorem, quantum theory and conceptions of the universe}},\ }\href@noop {} {\
  ,\ \bibinfo {pages} {69} (\bibinfo {year} {1989})}\BibitemShut {NoStop}%
\bibitem [{\citenamefont {Pan}\ and\ \citenamefont
  {Zeilinger}(1998)}]{pan1998greenberger}%
  \BibitemOpen
  \bibfield  {author} {\bibinfo {author} {\bibfnamefont {J.-W.}\ \bibnamefont
  {Pan}}\ and\ \bibinfo {author} {\bibfnamefont {A.}~\bibnamefont
  {Zeilinger}},\ }\bibfield  {title} {\bibinfo {title}
  {Greenberger-horne-zeilinger-state analyzer},\ }\href@noop {} {\bibfield
  {journal} {\bibinfo  {journal} {Physical Review A}\ }\textbf {\bibinfo
  {volume} {57}},\ \bibinfo {pages} {2208} (\bibinfo {year}
  {1998})}\BibitemShut {NoStop}%
\bibitem [{\citenamefont {Yang}\ \emph {et~al.}(2024)\citenamefont {Yang},
  \citenamefont {Mao}, \citenamefont {Chen}, \citenamefont {Dong},
  \citenamefont {Zhu}, \citenamefont {Wu},\ and\ \citenamefont
  {Li}}]{yang2024experimental}%
  \BibitemOpen
  \bibfield  {author} {\bibinfo {author} {\bibfnamefont {K.-X.}\ \bibnamefont
  {Yang}}, \bibinfo {author} {\bibfnamefont {Y.-L.}\ \bibnamefont {Mao}},
  \bibinfo {author} {\bibfnamefont {H.}~\bibnamefont {Chen}}, \bibinfo {author}
  {\bibfnamefont {X.}~\bibnamefont {Dong}}, \bibinfo {author} {\bibfnamefont
  {J.}~\bibnamefont {Zhu}}, \bibinfo {author} {\bibfnamefont {J.}~\bibnamefont
  {Wu}},\ and\ \bibinfo {author} {\bibfnamefont {Z.-D.}\ \bibnamefont {Li}},\
  }\bibfield  {title} {\bibinfo {title} {Experimental
  measurement-device-independent quantum conference key agreement},\
  }\href@noop {} {\bibfield  {journal} {\bibinfo  {journal} {Physical Review
  Letters}\ }\textbf {\bibinfo {volume} {133}},\ \bibinfo {pages} {210803}
  (\bibinfo {year} {2024})}\BibitemShut {NoStop}%
\bibitem [{\citenamefont {Du}\ \emph {et~al.}(2025)\citenamefont {Du},
  \citenamefont {Liu}, \citenamefont {Yang}, \citenamefont {Zheng},
  \citenamefont {Zhu},\ and\ \citenamefont {Ma}}]{du2025experimental}%
  \BibitemOpen
  \bibfield  {author} {\bibinfo {author} {\bibfnamefont {Y.}~\bibnamefont
  {Du}}, \bibinfo {author} {\bibfnamefont {Y.}~\bibnamefont {Liu}}, \bibinfo
  {author} {\bibfnamefont {C.}~\bibnamefont {Yang}}, \bibinfo {author}
  {\bibfnamefont {X.}~\bibnamefont {Zheng}}, \bibinfo {author} {\bibfnamefont
  {S.}~\bibnamefont {Zhu}},\ and\ \bibinfo {author} {\bibfnamefont {X.-s.}\
  \bibnamefont {Ma}},\ }\bibfield  {title} {\bibinfo {title} {Experimental
  measurement-device-independent quantum cryptographic conferencing},\
  }\href@noop {} {\bibfield  {journal} {\bibinfo  {journal} {Physical Review
  Letters}\ }\textbf {\bibinfo {volume} {134}},\ \bibinfo {pages} {040802}
  (\bibinfo {year} {2025})}\BibitemShut {NoStop}%
\bibitem [{Sup()}]{SupplementalMaterial}%
  \BibitemOpen
  \href@noop {} {}\bibinfo {note} {See Supplemental Material at [URL] for
  details of the protocol, pairing method, experimental implementation,
  experimental data, and data processing, which includes
  Refs.~[35--40].}\BibitemShut {Stop}%
\bibitem [{\citenamefont {Lo}\ \emph {et~al.}(2005)\citenamefont {Lo},
  \citenamefont {Ma},\ and\ \citenamefont {Chen}}]{lo2005decoy}%
  \BibitemOpen
  \bibfield  {author} {\bibinfo {author} {\bibfnamefont {H.-K.}\ \bibnamefont
  {Lo}}, \bibinfo {author} {\bibfnamefont {X.}~\bibnamefont {Ma}},\ and\
  \bibinfo {author} {\bibfnamefont {K.}~\bibnamefont {Chen}},\ }\bibfield
  {title} {\bibinfo {title} {Decoy state quantum key distribution},\
  }\href@noop {} {\bibfield  {journal} {\bibinfo  {journal} {Physical Review
  Letters}\ }\textbf {\bibinfo {volume} {94}},\ \bibinfo {pages} {230504}
  (\bibinfo {year} {2005})}\BibitemShut {NoStop}%
\bibitem [{\citenamefont {Wang}(2005)}]{wang2005beating}%
  \BibitemOpen
  \bibfield  {author} {\bibinfo {author} {\bibfnamefont {X.-B.}\ \bibnamefont
  {Wang}},\ }\bibfield  {title} {\bibinfo {title} {Beating the
  photon-number-splitting attack in practical quantum cryptography},\
  }\href@noop {} {\bibfield  {journal} {\bibinfo  {journal} {Physical Review
  Letters}\ }\textbf {\bibinfo {volume} {94}},\ \bibinfo {pages} {230503}
  (\bibinfo {year} {2005})}\BibitemShut {NoStop}%
\bibitem [{\citenamefont {Zhu}\ \emph {et~al.}(2023)\citenamefont {Zhu},
  \citenamefont {Huang}, \citenamefont {Liu}, \citenamefont {Zeng},
  \citenamefont {Zou}, \citenamefont {Dai}, \citenamefont {Tang}, \citenamefont
  {Li}, \citenamefont {You}, \citenamefont {Wang} \emph
  {et~al.}}]{zhu2023experimental}%
  \BibitemOpen
  \bibfield  {author} {\bibinfo {author} {\bibfnamefont {H.-T.}\ \bibnamefont
  {Zhu}}, \bibinfo {author} {\bibfnamefont {Y.}~\bibnamefont {Huang}}, \bibinfo
  {author} {\bibfnamefont {H.}~\bibnamefont {Liu}}, \bibinfo {author}
  {\bibfnamefont {P.}~\bibnamefont {Zeng}}, \bibinfo {author} {\bibfnamefont
  {M.}~\bibnamefont {Zou}}, \bibinfo {author} {\bibfnamefont {Y.}~\bibnamefont
  {Dai}}, \bibinfo {author} {\bibfnamefont {S.}~\bibnamefont {Tang}}, \bibinfo
  {author} {\bibfnamefont {H.}~\bibnamefont {Li}}, \bibinfo {author}
  {\bibfnamefont {L.}~\bibnamefont {You}}, \bibinfo {author} {\bibfnamefont
  {Z.}~\bibnamefont {Wang}}, \emph {et~al.},\ }\bibfield  {title} {\bibinfo
  {title} {Experimental mode-pairing measurement-device-independent quantum key
  distribution without global phase locking},\ }\href@noop {} {\bibfield
  {journal} {\bibinfo  {journal} {Physical Review Letters}\ }\textbf {\bibinfo
  {volume} {130}},\ \bibinfo {pages} {030801} (\bibinfo {year}
  {2023})}\BibitemShut {NoStop}%
\bibitem [{\citenamefont {Zhu}\ \emph {et~al.}(2026)\citenamefont {Zhu},
  \citenamefont {Huang}, \citenamefont {Rasmita}, \citenamefont {Ding},
  \citenamefont {Cai}, \citenamefont {Zhang}, \citenamefont {Ma},\ and\
  \citenamefont {Gao}}]{zhu2026frequency}%
  \BibitemOpen
  \bibfield  {author} {\bibinfo {author} {\bibfnamefont {H.-T.}\ \bibnamefont
  {Zhu}}, \bibinfo {author} {\bibfnamefont {Y.}~\bibnamefont {Huang}}, \bibinfo
  {author} {\bibfnamefont {A.}~\bibnamefont {Rasmita}}, \bibinfo {author}
  {\bibfnamefont {C.}~\bibnamefont {Ding}}, \bibinfo {author} {\bibfnamefont
  {X.}~\bibnamefont {Cai}}, \bibinfo {author} {\bibfnamefont {H.}~\bibnamefont
  {Zhang}}, \bibinfo {author} {\bibfnamefont {X.}~\bibnamefont {Ma}},\ and\
  \bibinfo {author} {\bibfnamefont {W.}~\bibnamefont {Gao}},\ }\bibfield
  {title} {\bibinfo {title} {Frequency-matching quantum key distribution},\
  }\href@noop {} {\bibfield  {journal} {\bibinfo  {journal} {Physical Review
  Applied}\ }\textbf {\bibinfo {volume} {25}},\ \bibinfo {pages} {034043}
  (\bibinfo {year} {2026})}\BibitemShut {NoStop}%
\bibitem [{\citenamefont {Shao}\ \emph {et~al.}(2025)\citenamefont {Shao},
  \citenamefont {Zhou}, \citenamefont {Lin}, \citenamefont {Minder},
  \citenamefont {Ge}, \citenamefont {Xie}, \citenamefont {Shen}, \citenamefont
  {Yan}, \citenamefont {Yin},\ and\ \citenamefont {Yuan}}]{shao2025high}%
  \BibitemOpen
  \bibfield  {author} {\bibinfo {author} {\bibfnamefont {S.-F.}\ \bibnamefont
  {Shao}}, \bibinfo {author} {\bibfnamefont {L.}~\bibnamefont {Zhou}}, \bibinfo
  {author} {\bibfnamefont {J.}~\bibnamefont {Lin}}, \bibinfo {author}
  {\bibfnamefont {M.}~\bibnamefont {Minder}}, \bibinfo {author} {\bibfnamefont
  {C.}~\bibnamefont {Ge}}, \bibinfo {author} {\bibfnamefont {Y.-M.}\
  \bibnamefont {Xie}}, \bibinfo {author} {\bibfnamefont {A.}~\bibnamefont
  {Shen}}, \bibinfo {author} {\bibfnamefont {Z.}~\bibnamefont {Yan}}, \bibinfo
  {author} {\bibfnamefont {H.-L.}\ \bibnamefont {Yin}},\ and\ \bibinfo {author}
  {\bibfnamefont {Z.}~\bibnamefont {Yuan}},\ }\bibfield  {title} {\bibinfo
  {title} {High-rate measurement-device-independent quantum communication
  without optical reference light},\ }\href@noop {} {\bibfield  {journal}
  {\bibinfo  {journal} {Physical Review X}\ }\textbf {\bibinfo {volume} {15}},\
  \bibinfo {pages} {021066} (\bibinfo {year} {2025})}\BibitemShut {NoStop}%
\bibitem [{\citenamefont {Zhang}\ \emph {et~al.}(2025)\citenamefont {Zhang},
  \citenamefont {Li}, \citenamefont {Pan}, \citenamefont {Lu}, \citenamefont
  {Li}, \citenamefont {Li}, \citenamefont {Huang}, \citenamefont {Ma},
  \citenamefont {Xu},\ and\ \citenamefont {Pan}}]{zhang2025experimental}%
  \BibitemOpen
  \bibfield  {author} {\bibinfo {author} {\bibfnamefont {L.}~\bibnamefont
  {Zhang}}, \bibinfo {author} {\bibfnamefont {W.}~\bibnamefont {Li}}, \bibinfo
  {author} {\bibfnamefont {J.}~\bibnamefont {Pan}}, \bibinfo {author}
  {\bibfnamefont {Y.}~\bibnamefont {Lu}}, \bibinfo {author} {\bibfnamefont
  {W.}~\bibnamefont {Li}}, \bibinfo {author} {\bibfnamefont {Z.-P.}\
  \bibnamefont {Li}}, \bibinfo {author} {\bibfnamefont {Y.}~\bibnamefont
  {Huang}}, \bibinfo {author} {\bibfnamefont {X.}~\bibnamefont {Ma}}, \bibinfo
  {author} {\bibfnamefont {F.}~\bibnamefont {Xu}},\ and\ \bibinfo {author}
  {\bibfnamefont {J.-W.}\ \bibnamefont {Pan}},\ }\bibfield  {title} {\bibinfo
  {title} {Experimental mode-pairing quantum key distribution surpassing the
  repeaterless bound},\ }\href@noop {} {\bibfield  {journal} {\bibinfo
  {journal} {Physical Review X}\ }\textbf {\bibinfo {volume} {15}},\ \bibinfo
  {pages} {021037} (\bibinfo {year} {2025})}\BibitemShut {NoStop}%
\bibitem [{\citenamefont {Lu}\ \emph {et~al.}(2025{\natexlab{b}})\citenamefont
  {Lu}, \citenamefont {Wang}, \citenamefont {Zhou}, \citenamefont {Fan},
  \citenamefont {Wang}, \citenamefont {Yin}, \citenamefont {Li}, \citenamefont
  {He}, \citenamefont {Wang}, \citenamefont {Chen} \emph
  {et~al.}}]{lu2025fully}%
  \BibitemOpen
  \bibfield  {author} {\bibinfo {author} {\bibfnamefont {F.-Y.}\ \bibnamefont
  {Lu}}, \bibinfo {author} {\bibfnamefont {Z.-H.}\ \bibnamefont {Wang}},
  \bibinfo {author} {\bibfnamefont {Y.}~\bibnamefont {Zhou}}, \bibinfo {author}
  {\bibfnamefont {Y.-X.}\ \bibnamefont {Fan}}, \bibinfo {author} {\bibfnamefont
  {S.}~\bibnamefont {Wang}}, \bibinfo {author} {\bibfnamefont {Z.-Q.}\
  \bibnamefont {Yin}}, \bibinfo {author} {\bibfnamefont {J.}~\bibnamefont
  {Li}}, \bibinfo {author} {\bibfnamefont {D.-Y.}\ \bibnamefont {He}}, \bibinfo
  {author} {\bibfnamefont {F.-X.}\ \bibnamefont {Wang}}, \bibinfo {author}
  {\bibfnamefont {W.}~\bibnamefont {Chen}}, \emph {et~al.},\ }\bibfield
  {title} {\bibinfo {title} {Fully heterogeneous prepare-and-measure quantum
  network for the next stage of quantum internet},\ }\href@noop {} {\bibfield
  {journal} {\bibinfo  {journal} {Nature Communications}\ }\textbf {\bibinfo
  {volume} {16}},\ \bibinfo {pages} {11487} (\bibinfo {year}
  {2025}{\natexlab{b}})}\BibitemShut {NoStop}%
\bibitem [{\citenamefont {Fang}\ \emph {et~al.}(2020)\citenamefont {Fang},
  \citenamefont {Zeng}, \citenamefont {Liu}, \citenamefont {Zou}, \citenamefont
  {Wu}, \citenamefont {Tang}, \citenamefont {Sheng}, \citenamefont {Xiang},
  \citenamefont {Zhang}, \citenamefont {Li} \emph
  {et~al.}}]{fang2020implementation}%
  \BibitemOpen
  \bibfield  {author} {\bibinfo {author} {\bibfnamefont {X.-T.}\ \bibnamefont
  {Fang}}, \bibinfo {author} {\bibfnamefont {P.}~\bibnamefont {Zeng}}, \bibinfo
  {author} {\bibfnamefont {H.}~\bibnamefont {Liu}}, \bibinfo {author}
  {\bibfnamefont {M.}~\bibnamefont {Zou}}, \bibinfo {author} {\bibfnamefont
  {W.}~\bibnamefont {Wu}}, \bibinfo {author} {\bibfnamefont {Y.-L.}\
  \bibnamefont {Tang}}, \bibinfo {author} {\bibfnamefont {Y.-J.}\ \bibnamefont
  {Sheng}}, \bibinfo {author} {\bibfnamefont {Y.}~\bibnamefont {Xiang}},
  \bibinfo {author} {\bibfnamefont {W.}~\bibnamefont {Zhang}}, \bibinfo
  {author} {\bibfnamefont {H.}~\bibnamefont {Li}}, \emph {et~al.},\ }\bibfield
  {title} {\bibinfo {title} {Implementation of quantum key distribution
  surpassing the linear rate-transmittance bound},\ }\href@noop {} {\bibfield
  {journal} {\bibinfo  {journal} {Nature Photonics}\ }\textbf {\bibinfo
  {volume} {14}},\ \bibinfo {pages} {422} (\bibinfo {year} {2020})}\BibitemShut
  {NoStop}%
\bibitem [{\citenamefont {Liu}\ \emph {et~al.}(2023)\citenamefont {Liu},
  \citenamefont {Zhang}, \citenamefont {Jiang}, \citenamefont {Chen},
  \citenamefont {Zhang}, \citenamefont {Pan}, \citenamefont {Ma}, \citenamefont
  {Dong}, \citenamefont {Xiong}, \citenamefont {Zhang} \emph
  {et~al.}}]{liu2023experimental}%
  \BibitemOpen
  \bibfield  {author} {\bibinfo {author} {\bibfnamefont {Y.}~\bibnamefont
  {Liu}}, \bibinfo {author} {\bibfnamefont {W.-J.}\ \bibnamefont {Zhang}},
  \bibinfo {author} {\bibfnamefont {C.}~\bibnamefont {Jiang}}, \bibinfo
  {author} {\bibfnamefont {J.-P.}\ \bibnamefont {Chen}}, \bibinfo {author}
  {\bibfnamefont {C.}~\bibnamefont {Zhang}}, \bibinfo {author} {\bibfnamefont
  {W.-X.}\ \bibnamefont {Pan}}, \bibinfo {author} {\bibfnamefont
  {D.}~\bibnamefont {Ma}}, \bibinfo {author} {\bibfnamefont {H.}~\bibnamefont
  {Dong}}, \bibinfo {author} {\bibfnamefont {J.-M.}\ \bibnamefont {Xiong}},
  \bibinfo {author} {\bibfnamefont {C.-J.}\ \bibnamefont {Zhang}}, \emph
  {et~al.},\ }\bibfield  {title} {\bibinfo {title} {Experimental twin-field
  quantum key distribution over 1000 km fiber distance},\ }\href@noop {}
  {\bibfield  {journal} {\bibinfo  {journal} {Physical Review Letters}\
  }\textbf {\bibinfo {volume} {130}},\ \bibinfo {pages} {210801} (\bibinfo
  {year} {2023})}\BibitemShut {NoStop}%
\bibitem [{\citenamefont {Wang}\ \emph {et~al.}(2022)\citenamefont {Wang},
  \citenamefont {Yin}, \citenamefont {He}, \citenamefont {Chen}, \citenamefont
  {Wang}, \citenamefont {Ye}, \citenamefont {Zhou}, \citenamefont {Fan-Yuan},
  \citenamefont {Wang}, \citenamefont {Chen} \emph {et~al.}}]{wang2022twin}%
  \BibitemOpen
  \bibfield  {author} {\bibinfo {author} {\bibfnamefont {S.}~\bibnamefont
  {Wang}}, \bibinfo {author} {\bibfnamefont {Z.-Q.}\ \bibnamefont {Yin}},
  \bibinfo {author} {\bibfnamefont {D.-Y.}\ \bibnamefont {He}}, \bibinfo
  {author} {\bibfnamefont {W.}~\bibnamefont {Chen}}, \bibinfo {author}
  {\bibfnamefont {R.-Q.}\ \bibnamefont {Wang}}, \bibinfo {author}
  {\bibfnamefont {P.}~\bibnamefont {Ye}}, \bibinfo {author} {\bibfnamefont
  {Y.}~\bibnamefont {Zhou}}, \bibinfo {author} {\bibfnamefont {G.-J.}\
  \bibnamefont {Fan-Yuan}}, \bibinfo {author} {\bibfnamefont {F.-X.}\
  \bibnamefont {Wang}}, \bibinfo {author} {\bibfnamefont {W.}~\bibnamefont
  {Chen}}, \emph {et~al.},\ }\bibfield  {title} {\bibinfo {title} {Twin-field
  quantum key distribution over 830-km fibre},\ }\href@noop {} {\bibfield
  {journal} {\bibinfo  {journal} {Nature photonics}\ }\textbf {\bibinfo
  {volume} {16}},\ \bibinfo {pages} {154} (\bibinfo {year} {2022})}\BibitemShut
  {NoStop}%
\bibitem [{\citenamefont {Chen}\ \emph {et~al.}(2020)\citenamefont {Chen},
  \citenamefont {Zhang}, \citenamefont {Liu}, \citenamefont {Jiang},
  \citenamefont {Zhang}, \citenamefont {Hu}, \citenamefont {Guan},
  \citenamefont {Yu}, \citenamefont {Xu}, \citenamefont {Lin} \emph
  {et~al.}}]{chen2020sending}%
  \BibitemOpen
  \bibfield  {author} {\bibinfo {author} {\bibfnamefont {J.-P.}\ \bibnamefont
  {Chen}}, \bibinfo {author} {\bibfnamefont {C.}~\bibnamefont {Zhang}},
  \bibinfo {author} {\bibfnamefont {Y.}~\bibnamefont {Liu}}, \bibinfo {author}
  {\bibfnamefont {C.}~\bibnamefont {Jiang}}, \bibinfo {author} {\bibfnamefont
  {W.}~\bibnamefont {Zhang}}, \bibinfo {author} {\bibfnamefont {X.-L.}\
  \bibnamefont {Hu}}, \bibinfo {author} {\bibfnamefont {J.-Y.}\ \bibnamefont
  {Guan}}, \bibinfo {author} {\bibfnamefont {Z.-W.}\ \bibnamefont {Yu}},
  \bibinfo {author} {\bibfnamefont {H.}~\bibnamefont {Xu}}, \bibinfo {author}
  {\bibfnamefont {J.}~\bibnamefont {Lin}}, \emph {et~al.},\ }\bibfield  {title}
  {\bibinfo {title} {Sending-or-not-sending with independent lasers: Secure
  twin-field quantum key distribution over 509 km},\ }\href@noop {} {\bibfield
  {journal} {\bibinfo  {journal} {Physical review letters}\ }\textbf {\bibinfo
  {volume} {124}},\ \bibinfo {pages} {070501} (\bibinfo {year}
  {2020})}\BibitemShut {NoStop}%
\bibitem [{\citenamefont {Chen}\ \emph {et~al.}(2024)\citenamefont {Chen},
  \citenamefont {Zhou}, \citenamefont {Zhang}, \citenamefont {Jiang},
  \citenamefont {Chen}, \citenamefont {Huang}, \citenamefont {Li},
  \citenamefont {You}, \citenamefont {Wang}, \citenamefont {Liu} \emph
  {et~al.}}]{chen2024twin}%
  \BibitemOpen
  \bibfield  {author} {\bibinfo {author} {\bibfnamefont {J.-P.}\ \bibnamefont
  {Chen}}, \bibinfo {author} {\bibfnamefont {F.}~\bibnamefont {Zhou}}, \bibinfo
  {author} {\bibfnamefont {C.}~\bibnamefont {Zhang}}, \bibinfo {author}
  {\bibfnamefont {C.}~\bibnamefont {Jiang}}, \bibinfo {author} {\bibfnamefont
  {F.-X.}\ \bibnamefont {Chen}}, \bibinfo {author} {\bibfnamefont
  {J.}~\bibnamefont {Huang}}, \bibinfo {author} {\bibfnamefont
  {H.}~\bibnamefont {Li}}, \bibinfo {author} {\bibfnamefont {L.-X.}\
  \bibnamefont {You}}, \bibinfo {author} {\bibfnamefont {X.-B.}\ \bibnamefont
  {Wang}}, \bibinfo {author} {\bibfnamefont {Y.}~\bibnamefont {Liu}}, \emph
  {et~al.},\ }\bibfield  {title} {\bibinfo {title} {Twin-field quantum key
  distribution with local frequency reference},\ }\href@noop {} {\bibfield
  {journal} {\bibinfo  {journal} {Physical Review Letters}\ }\textbf {\bibinfo
  {volume} {132}},\ \bibinfo {pages} {260802} (\bibinfo {year}
  {2024})}\BibitemShut {NoStop}%
\bibitem [{\citenamefont {Zhang}\ \emph {et~al.}(2017)\citenamefont {Zhang},
  \citenamefont {Zhao}, \citenamefont {Razavi},\ and\ \citenamefont
  {Ma}}]{zhang2017improved}%
  \BibitemOpen
  \bibfield  {author} {\bibinfo {author} {\bibfnamefont {Z.}~\bibnamefont
  {Zhang}}, \bibinfo {author} {\bibfnamefont {Q.}~\bibnamefont {Zhao}},
  \bibinfo {author} {\bibfnamefont {M.}~\bibnamefont {Razavi}},\ and\ \bibinfo
  {author} {\bibfnamefont {X.}~\bibnamefont {Ma}},\ }\bibfield  {title}
  {\bibinfo {title} {Improved key-rate bounds for practical decoy-state
  quantum-key-distribution systems},\ }\href@noop {} {\bibfield  {journal}
  {\bibinfo  {journal} {Physical Review A}\ }\textbf {\bibinfo {volume} {95}},\
  \bibinfo {pages} {012333} (\bibinfo {year} {2017})}\BibitemShut {NoStop}%
\end{thebibliography}%

\end{document}